\documentclass[prb,twocolumn,superscriptaddress,floatfix,letterpaper,showpacs]{revtex4}
\usepackage{graphicx,amssymb,bm,amsfonts,natbib,amsmath,graphics,color}

\usepackage[bookmarks=false,pdfstartview=FitH,colorlinks=true,citecolor=blue,linkcolor=blue]{hyperref}
\begin{document}

\title {Bilayer splitting and c-axis coupling in CMR bilayer manganites}

\author{C. Jozwiak} \affiliation{Department of Physics, University of California Berkeley, California 94720, USA} \affiliation{Materials Sciences Division, Lawrence Berkeley National Laboratory, Berkeley, California 94720, USA}
\author{J. Graf} \affiliation{Materials Sciences Division, Lawrence Berkeley National Laboratory, Berkeley, California 94720, USA}
\author{S. Y. Zhou} \affiliation{Department of Physics, University of California Berkeley, California 94720, USA} \affiliation{Materials Sciences Division, Lawrence Berkeley National Laboratory, Berkeley, California 94720, USA}
\author{A. Bostwick} \affiliation{Advanced Light Source, Lawrence Berkeley National Laboratory, Berkeley, California 94720, USA}
\author{Eli. Rotenberg} \affiliation{Advanced Light Source, Lawrence Berkeley National Laboratory, Berkeley, California 94720, USA}
\author{H. Zheng} \affiliation{Materials Science Division, Argonne National Laboratory, Argonne, Illinois, 60439, USA}
\author{J. F. Mitchell} \affiliation{Materials Science Division, Argonne National Laboratory, Argonne, Illinois, 60439, USA}
\author{A. Lanzara} \email{alanzara@lbl.gov} \affiliation{Department of Physics, University of California Berkeley, California 94720, USA} \affiliation{Materials Sciences Division, Lawrence Berkeley National Laboratory, Berkeley, California 94720, USA} 

\date {\today}

\begin{abstract}
By performing angle-resolved photoemission spectroscopy of the bilayer colossal magnetoresistive (CMR) manganite, La$_{2-2x}$Sr$_{1+2x}$Mn$_{2}$O$_{7}$, we provide the complete mapping of the Fermi level spectral weight topology.
Clear and unambiguous bilayer splitting of the in-plane 3\textit{d}$_{x^2-y^2}$ band, mapped throughout the Brillouin zone, and the full mapping of the 3\textit{d}$_{3z^2-r^2}$ band are reported.
Peculiar doping and temperature dependencies of these bands imply that as transition from the ferromagnetic metallic phase approaches, either as a function of doping or temperature, coherence along the c-axis between planes within the bilayer is lost, resulting in reduced interplane coupling.
These results suggest that interplane coupling plays a large role in the CMR transition.
\end{abstract}

\pacs{75.47.Gk, 71.30.+h, 71.18.+y, 79.60.-i}

\keywords{manganites; strongly correlated electron systems; Colossal Magnetoresistance; CMR; Bilayer Splitting}
\maketitle

\section{\label{sec:Intro}INTRODUCTION}

The bilayer colossal magnetoresistive (CMR) manganites, La$_{2-2x}$Sr$_{1+2x}$Mn$_{2}$O$_{7}$, are characterized by a layered structure of pairs of MnO$_2$ planes separated by insulating (La,Sr) slabs.
They display such large structural and transport anisotropies \cite{Moritomo1996} that their electronic structure is assumed to be largely 2-dimensional.
The single layered analogs, however, do not exhibit the metallic or ferromagnetic phases characteristic of the CMR compounds, suggesting that c-axis coupling between the MnO$_2$ planes, \textit{within} the bilayer, is relevant to CMR physics.
This may be supported by the observations of doping dependent decreases of the Jahn-Teller distortions,\cite{Kimura1998,Kubota2000} c-axis directed orbital occupancies,\cite{Koizumi2001,Koizumi2005} and the inter-plane exchange coupling within the bilayer.\cite{Perring2001,Hirota2002}
Because bulk c-axis transport is dominated by the weak coupling \textit{between} bilayers, it is blind to the interlayer coupling \textit{within} the bilayers.  
Angle-resolved photoemission spectroscopy (ARPES), on the other hand, can be directly sensitive to these out-of-plane contributions to the electronic structure.
This may explain why transport shows little difference upon changing doping from \textit{x}=0.36 to 0.40, while substantial differences in the electronic structure have been reported (e.g. presence of coherent quasiparticles throughout momentum space in the former \cite{Sun2006,Jong2007} and momentum space localized coherent quasiparticles in the latter \cite{Mannella2005}).
Although ARPES has been successfully used to study the bilayer manganites,\cite{Dessau1998,Saitoh2000,ChuangCMR2001,Mannella2005,Sun2006,Jong2007,Sun2008} so far no ARPES work has been directed at understanding the out-of-plane components of the electronic structure and their relation to CMR.

In this paper, we present an ARPES study directed at addressing this important issue. 
We present data for two dopings, 0.36 and 0.40, corresponding to the doping with the maximum \textit{T}$_C$ of low temperature ferromagnetic ordering and the doping with slight antiferromagnetic canting between planes within the bilayer,\cite{Osborn1998,Kubota1999} respectively.
We report the complete mapping of the Fermi level spectral weight topology of the bilayer split in-plane bands and the out-of-plane band.
We find that the bilayer splitting, which is a sign of coherent coupling between the intra-bilayer planes, is dramatically reduced for the higher doping and above \textit{T}$_C$ for the lower doping.
We also find that the out-of-plane band, here well resolved throughout momentum space, loses definition with both increasing doping and temperature through \textit{T}$_C$.
These results signify a change in the electronic structure through a reduction of the inter-plane (intra-bilayer) coupling as the low temperature ferromagnetic metallic (FM) phase approaches its borders in the phase diagram with increasing doping or temperature, and therefore they identify important characteristics of the electronic structure which bulk measurements are blind to.

\section{\label{sec:Exp}EXPERIMENTAL DETAILS}

The ARPES experiments were performed at beamline 7.0.1.2 of the Advanced Light Source, Berkeley, on high-quality single crystals of La$_{2-2x}$Sr$_{1+2x}$Mn$_{2}$O$_{7}$ (x=0.36,0.40), grown with the floating zone method.\cite{Mitchell1997}
Samples were cleaved along the \textit{ab} plane in vacuum of 3 x 10$^{-11}$ torr.
Samples of \textit{x}=0.40 were cleaved at 160K with data taken at successively lower temperatures to 20K, ensuring that the fewer features seen at 160K are not due to surface degradation.
Samples of \textit{x}=0.36 were cleaved at 20K with data taken at successively higher temperatures to 160K, and then cycled back to 25K.  The final 25K data are identical to the initial 20K data, again confirming that the temperature dependence observed is not due to surface aging.
All data shown here were taken with 150~eV photons with polarization $\sim$60$^{\circ}$ out-of-plane and the in-plane component along the $\Gamma$M direction. 
The combined instrumental energy resolution was less than 30~meV, and the momentum resolution was better than 0.025~$\pi$/a.

ARPES is a surface sensitive technique, and so must be performed on high-quality surfaces representative of the bulk material for data to be interpreted as intrinsic to the bulk.
\textit{In-situ} cleaving of bilayer manganite crystals is known to give extremely flat\cite{Ronnow2006} mirror-like surfaces which give clear low-energy electron diffraction (LEED) patterns, without superlattice spots, indicating high quality surfaces free of reconstruction.\cite{Dessau1998,ChuangCMR2001,Jong2007}
Accordingly, a large number of ARPES works have been performed on such cleaved surfaces within the common assumption of bulk representation.\cite{Dessau1998,Saitoh2000,ChuangCMR2001,Mannella2005,Sun2006,Jong2007,Sun2008}
The current work has been performed on a large number of separate cleaves to ensure that the features (or lack of features) presented in both dopings are not due to poorly cleaved surfaces.

\section{\label{sec:Results}RESULTS AND DISCUSSION}

\begin{figure} \includegraphics[width=8.5cm]{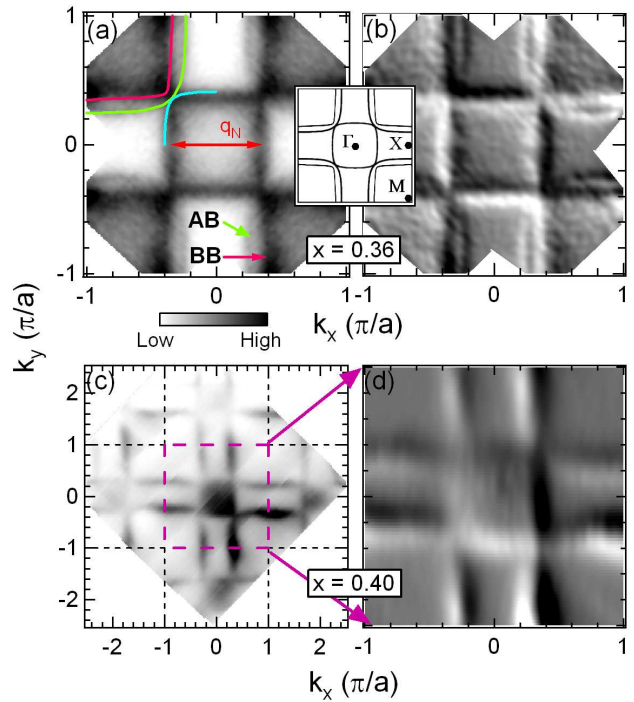}
\caption{\label{fig:FS}(Color online). ARPES derived FS integrated from -90meV to E$_F$ at 20K.  (a) FS of \textit{x}=0.36 showing the BB and AB IP bands and the OP electron pocket at $\Gamma$.  Data from one quadrant is symmetrized to the other three.  Calculated FS pieces from Refs.~\onlinecite{Huang2000}~and~\onlinecite{Mannella2005} are overplotted in the upper left.  (b) First derivative of the intensity with respect to \textit{k}$_x$ and \textit{k}$_y$ of the data in (a) to highlight features.  The inset shows a representative FS. (c)  FS of \textit{x}=0.40 covering multiple Brillouin zones (unsymmetrized).  (d) First derivative of (c), only showing the first zone for direct comparisson with (b).}
\end{figure}

Figure~\ref{fig:FS} shows the measured Fermi surface (FS) at 20K and its first derivative for \textit{x}=0.36 (panels a-b) and \textit{x}=0.40 (panels c-d), both exhibiting CMR. 
The FS here is defined as the maximal intensity contours at the Fermi level, and as described in Ref.~\onlinecite{Mannella2005}, does not strictly distinguish a Fermi surface of sharp quasiparticles from a `ghost-like' surface.
The expected FS is shown in the inset and is formed by three pieces.\cite{Dessau1998,Huang2000}
The first is a square electron pocket at $\Gamma$ with mainly out-of-plane (OP) 3\textit{d}$_{3z^2-r^2}$ orbital character. 
The other pieces are concentric hole pockets around each M point, formed by a mainly in-plane (IP) 3\textit{d}$_{x^2-y^2}$ derived band split into two bonding (BB) and antibonding (AB) bands.
This splitting results from coupling of the two identical MnO$_2$ layers in the unit cell, and is known as bilayer splitting (BLS).
The calculated FS pieces from the electron pocket (from Ref.~\onlinecite{Huang2000}) and the BB (from Ref.~\onlinecite{Mannella2005}) and AB (from Refs.~\onlinecite{Huang2000}~and~\onlinecite{Mannella2005}) hole pockets are overplotted in one quadrant of the data in Fig.~\ref{fig:FS}(a).
Previous works have successfully mapped only one of the hole pockets (the two pockets were resolved separately with two photon energies for \textit{x}=0.38,0.36,\cite{Sun2006} but without full \textit{k}-space mappings), and have not clearly resolved the electron pocket.\cite{Dessau1998, ChuangCMR2001, Mannella2005} 
Higher photon energy and out-of-plane polarization allowed us to resolve the full FS topology of all three pockets simultaneously for \textit{x}=0.36.
We will focus on the IP bands first, and then address the OP band in detail with Figs.~\ref{fig:GX}~and~\ref{fig:OP}.

The \textit{x}=0.36 data resolves all three components of the expected FS and shows good agreement with theory, \cite{Mannella2005,Huang2000} particularly in regards to the AB (BB) being more ``circular'' (``square'') and the BLS being largest along the XM direction.\cite{Mannella2005}
Indeed, the present measurement of the full topology of the BB piece suggests better agreement with the calculation shown in Ref.~\onlinecite{Mannella2005} than that given in Ref.~\onlinecite{Huang2000}.
The measured splitting appears slightly larger than the calculation (the separation along XM between the corresponding momentum distribution curve (MDC) peaks at \textit{E}$_F$ is $\sim$35\% larger than the calculated band positions at \textit{E}$_F$).
The calculation is for \textit{x}=0.40, however, and the splitting is expected to increase with decreasing doping,\cite{Huang2000} in line with our observation.
In contrast, for \textit{x}=0.40 measured with identical experimental conditions, we could barely resolve the electron pocket and no evidence of BLS is observed anywhere in k-space, resulting in only single hole pockets in the FS maps [Fig.~\ref{fig:FS}(c)~and~\ref{fig:FS}(d)].
Although it is unlikely that the ARPES matrix elements are doping dependent, we searched a range of photon energies from 80 to 200~eV (not shown) and a momentum region up to the third Brillouin zone [Fig.~\ref{fig:FS}(c)] to ensure these are not matrix element effects.
This doping dependence is unlikely to be due to different surface qualities, as both dopings gave equally high quality surfaces by eye and a large number of experimental runs on separate cleaves gave consistent results.

\begin{figure} \includegraphics[width=8.5cm]{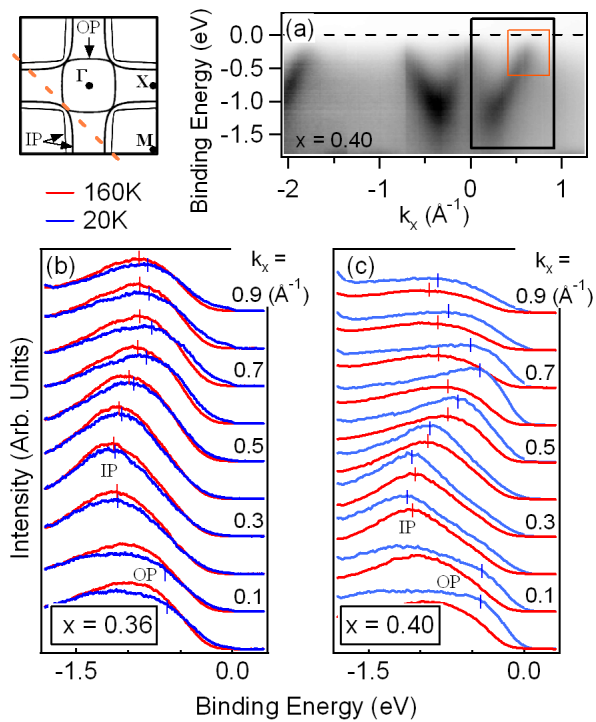}
\caption{\label{fig:OGM}(Color online). (a) Photoemission intensity map as a function of binding energy and momentum along a cut parallel to the $\Gamma$M direction, defined by the dashed orange line in the FS inset, for \textit{x}=0.40 at 20K.  Here, $k=0$ is defined as the intersection with the perpendicular line passing through the $\Gamma$ and M points.  (b) EDCs at the labeled momentum positions along this cut for \textit{x}=0.36 at 20K (blue) and 160K (red).  Vertical dashes are meant as guides-to-the-eyes of the approximate peak locations of the bands identified as labeled.  The momentum and energy range of these EDCs is illustrated by the black box in panel (a).  The smaller, orange box in panel (a) illustrates the range of data included in Fig.~\ref{fig:BLSX}.  (c) The corresponding EDCs for \textit{x}=0.40.
}\end{figure}

Figure~\ref{fig:OGM} displays ARPES data along the cut in momentum space defined by the dashed line in the FS inset to the figure; note that this cut passes through the point on the FS along the Brillouin zone boundary where the BLS should be the largest.
Panel (a) shows an image plot of the ARPES intensity along this cut through slightly more than one zone for \textit{x}=0.40.
This cut passes through a somewhat complex region of the bandstructure.
The visible band dispersing from near \textit{E}$_F$ at $k=0$ downwards toward $k=\pm 0.2$~\AA$^{-1}$ contains weight from all three bands as they nearly intersect at the FS at this momentum location.
The clear band dispersing upwards from $k=\pm 0.3$~\AA$^{-1}$ toward \textit{E}$_F$ near $k=\pm 0.7$~\AA$^{-1}$ contains weight from just the IP bands which should cross E$_F$ as distinct bands causing the two separate BB and AB pieces of FS as shown in the inset.
Energy distribution curves (EDCs: intensity as a function of energy at different momentum values) taken from such maps for both \textit{x}=0.36 and \textit{x}=0.40 are shown in Figs.~\ref{fig:OGM}(b)~and~\ref{fig:OGM}(c), respectively.
Several key features are common to both plots.
The described dispersions contain only single, broad incoherent peaks which are too broad to resolve any BLS and do not disperse all the way to \textit{E}$_F$ at the expected \textit{k}$_F$ locations, even though the momentum topology of the ARPES-derived FS maps in Fig.~\ref{fig:FS} clearly follows the expected FS.
Additionally,  the EDC peaks which are closest to \textit{E}$_F$ (near the expected \textit{k}$_F$ locations) disperse significantly closer to \textit{E}$_F$ at temperatures below \textit{T}$_C$ compared with above.
These characteristics agree with previous works\cite{Dessau1998,Saitoh2000} in which they are discussed in detail.
Interestingly, none of the EDCs from either doping show any evidence of sharp quasiparticle peaks which have been the focus of more recent ARPES works on these samples.\cite{Mannella2005,Sun2006,Jong2007,Sun2008}
The lack of these sharp peaks here does not mean that coherent quasiparticles are not present in the studied samples, but only that the ARPES matrix elements resulting from the present experimental parameters are not favorable for their observation.

As a brief aside, it is important to note that photoemission matrix element effects can greatly change ARPES results dependent on initial state symmetry, experimental geometry, and photon energy and polarization.
In fact, it is clear that use of differing photon energy dependencies of the AB and BB bands in both bilayer superconducting cuprates \cite{Feng2001,ChuangHTC2001} and the present manganites \cite{Sun2006} can allow one of the bands to be better exposed at the expense of the other.
As another example of the impact of matrix element modulation, the recent observation of coherent quasiparticles in \textit{x}=0.40 manganites was made using photon polarization which was found to give worse exposure of spectral weight near $\Gamma$.\cite{Mannella2005}
These illustrate the general tendency of matrix elements to provide give-and-take scenarios, where one must choose which features to enhance at the expense of others.
In the present data, geometry and photon energies were chosen to provide simultaneous access to the momentum distribution of spectral weight for all of the near \textit{E}$_F$ bands.
This has provided a straightforward way to investigate the doping and temperature dependence of aspects such as BLS and the OP electron pocket, not previously directly studied.
These benefits apparently come at the expense of resolving the quasiparticles (sharp, near \textit{E}$_F$ peaks in the EDCs) of much recent attention, \cite{Mannella2005,Sun2006,Jong2007} disallowing analysis techniques such as extraction of electron-phonon coupling parameters.\cite{Mannella2005,Sun2006}

\begin{figure} \includegraphics[width=7cm]{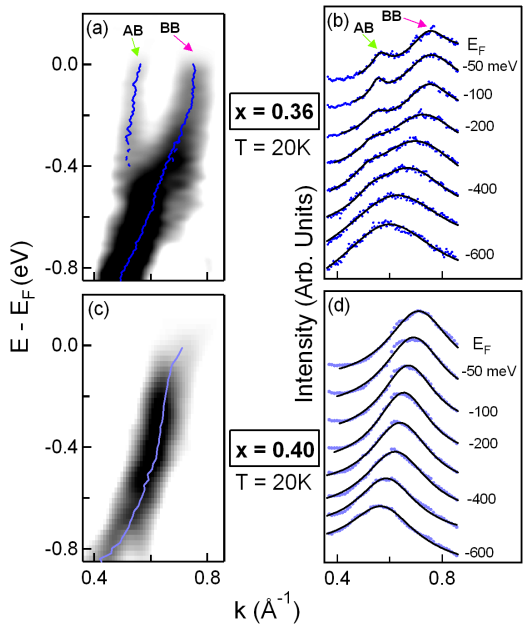}
\caption{\label{fig:BLSX}(Color online).  ARPES data at 20K along the cut shown by the dashed line in the FS inset of Fig.~\ref{fig:OGM} through the energy and momentum range shown by the small orange box in Fig.~\ref{fig:OGM}(a).  BLS is clearly seen in \textit{x}=0.36 and not observed in \textit{x}=0.40.  (a) Second derivative ($d^2/dk^2$) of ARPES intensity of binding energy versus momentum for \textit{x}=0.36.  The MDC peak dispersion is overplotted in blue (dotted portions mark energy range in which MDC extracted dispersion is less trustworthy as two separate peaks become more difficult to resolve).  (b) MDCs of raw data, normalized to equal peak intensity after constant background subtraction.  Lorentzian fits overplotted as black curves.  (c) Similar second derivative ARPES map for \textit{x}=0.40.  (d) Corresponding MDCs normalized to equal peak intensity after constant background subtraction. Lorenztian fits overplotted as black curves.}
\end{figure}

To examine the BLS in detail, Fig.~\ref{fig:BLSX} displays ARPES data for the same cut shown in Fig.~\ref{fig:OGM}, focused on the region near \textit{E}$_F$ [the orange box in Fig.~\ref{fig:OGM}(a)], where the BLS is expected to be the largest.
Despite the lack of measured quasiparticle peaks in the EDCs, significant information is still present in the momentum distribution of the ARPES intensity.
In particular, for the \textit{x}=0.36 sample, two separate bands crossing the Fermi level (panel a) and corresponding two peaks in the raw MDCs (panel b) can be clearly resolved between -400meV and \textit{E}$_F$.
In contrast, the \textit{x}=0.40 doping has a single band crossing the Fermi level with a single peak in the MDC through the full energy range (panels c~and~d).
Although it is possible that BLS is present in the \textit{x}=0.40 samples at magnitudes not resolvable with our present experimental resolution, it can be concluded that BLS in \textit{x}=0.40 is significantly reduced compared to \textit{x}=0.36, as the widths of the single peaks in the \textit{x}=0.40 MDCs are much less than the total widths of the double peaks in the \textit{x}=0.36 MDCs.
Note that the small glitch present in the MDCs of Fig.~\ref{fig:BLSX}(d) near 0.62~\AA$^{-1}$ is only an artifact from combining two separate windows of momentum range after data acquisition.
This is supported by its location (the glitch appears at the border between the two acquired data ranges) and by its non-dispersive nature, contrary to the dispersing BB and AB peaks observed in panel (b) for the 0.36 sample.

Figures~\ref{fig:FS}-\ref{fig:BLSX} show a decrease in BLS, suggesting a loss of coherence along the out-of plane direction as the doping increases.
BLS is present if there is coherence along the out-of-plane direction at least between two planes within the bilayer.\cite{ChuangHTC2004,Kaminski2003}.
If correlations preclude coherent hopping between these planes, BLS will be absent.\cite{Kaminski2003}
Possible explanations for the disappearence of BLS at higher doping include the decrease of the out-of-plane orbital occupancy \cite{Koizumi2001,Koizumi2005} which is likely related to the decrease of the Jahn-Teller distortions.\cite{Kimura1998,Kubota2000,Jackeli2002}
These trends may be the source of the differences in electronic structure presented here, as well as the root of the differences previously reported for these two dopings.\cite{Mannella2005,Sun2006}
The concomitant decrease in interplane exchange coupling \cite{Perring2001,Hirota2002} and slight canting of spin alignment between the planes within the bilayer with increasing doping \cite{Osborn1998,Kubota1999} are also signs of decreased coupling and may be involved.

\begin{figure}[!t] \includegraphics[width=8cm]{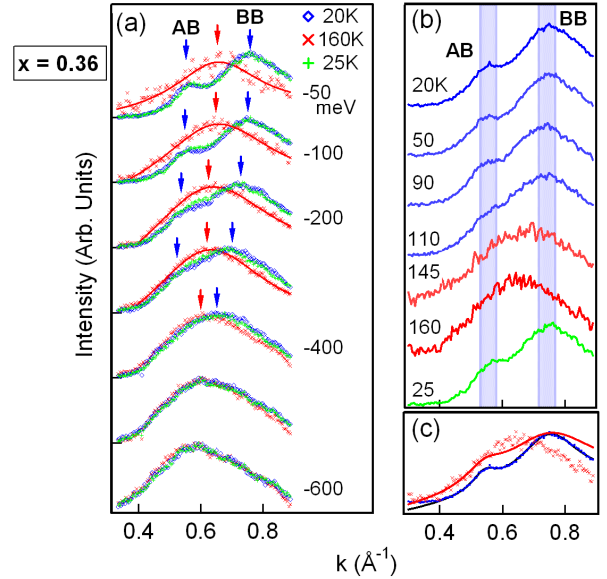}
\caption{\label{fig:BLST}(Color online). Temperature dependence of the \textit{x}=0.36 BLS for the cut in Fig.~\ref{fig:BLSX} (along the dashed line in the Fig.~\ref{fig:OGM} FS inset).  (a) MDCs (normalized to equal peak intensity after constant background subtraction) of ARPES intensity maps of binding energy vs.~momentum at 20 (blue), 160 (red), and after cycling back to 25 K (green).  Lorentzian fits of 160K data are overplotted in solid red.  Peak positions for 20 K (160 K) fits are shown with blue (red) arrows.  (b) MDCs at -100~meV shown for a range of temperatures.  Experiment proceeded from top to bottom.  Blue stripes highlight the AB and BB peaks below $T_C$.  (c) MDC at -100~meV of 20K data, artificially energy-broadened equivalent to 160 K (blue dots), with two-peak Lorentzian fit (solid blue).  This fit, with peak widths broadened by 1.5 times is in solid red.  These broadenings cannot explain the lineshape of the actual 160 K MDC (red dots).
}\end{figure}

The temperature dependence of BLS in \textit{x}=0.36 across the FM to paramagnetic insulating (PI) transition (\textit{T}$_C$$\sim$132K) is shown in Fig.~\ref{fig:BLST}.
MDCs above and below \textit{T}$_C$ are compared in Fig.~\ref{fig:BLST}(a).
At binding energies larger than -400meV the spectra agree perfectly. 
Surprisingly, where the BLS is observed at low temperature, -400meV $\leq$ \textit{E} $\le$ \textit{E}$_F$, a dramatic change of lineshape occurs and only a single broad peak, located between the AB and BB peaks, is observed above \textit{T}$_C$.
Figure~\ref{fig:BLST}(b) shows that this change occurs rather suddenly at \textit{T}$_C$, and is fully reversible.
Broadening the 20K spectra in energy to accommodate thermal broadening, as well as broadening the resulting MDCs cannot explain the 160K data, as shown in panel (c).

This loss in BLS may be another manifestation of loss of electronic coherence along the c-axis within the bilayers.
Other ARPES works have also found signatures of coherence loss above \textit{T}$_C$ through the disappearance of quasiparticles,\cite{Mannella2005,Mannella2007} interpreted as a crossover from a liquid of coherently moving Jahn-Teller related large polarons in the FM state, to thermally hopping small polarons in the PI state.\cite{Millis1996,Lanzara1998,Vasiliu1999,Mannella2005,Mannella2007} 
The present results, together with the lack of a FM phase in the single-layer compounds,\cite{Moritomo1996} suggest that 
coherent coupling between intrabilayer planes may play a role in the FM phase and CMR transition in the bilayer samples.
The loss of coupling may help swing the balance toward the incoherent small polaron PI state. 
As doping is raised to 0.48, the bilayers have full \textit{A}-type antiferromagnetic order between the intrabilayer planes at low temperature (suggesting the planes are decoupled), and the material's groundstate is insulating,\cite{Li2003} in line with this picture.
One should note, however, that samples of even higher doping ($x>0.54$) are also \textit{A}-type antiferromagnetic, and yet are metallic at low temperature\cite{Badica2004} and therefore not directly in line with this picture.
Indeed, samples in this range of doping are the most metallic of any layered manganites,\cite{Li2006} which shows that interplane, intra-bilayer ferromagnetic coupling is not generally required for metallicity. 
However, these fascinating samples inhabit an extremely sensitive region of the phase diagram, densely packed with exotic behavior, and the differing balance of competing orders may entail varied or further explanations.
As discussed in earlier works, a striking anisotropy in behaviors between doping regions of $x<0.5$ and $x>0.5$ may require additional consideration of effects such as orbital degeneracy and stronger Coulomb interactions at the higher doping levels.\cite{Badica2004,Brink1999}

It is tempting to relate this proposed effect to the dimensional crossovers recently studied in the cobaltates \cite{Valla2002} and ruthenates.\cite{Wang2004}
However, these effects are fundamentally different as they involve coherence crossovers from two dimensions to full bulk three dimensions which are visible in bulk transport measurements.
The present results are unique in that they show a change in c-axis coherence restricted \textit{within} the bilayers that is not visible to transport measurements that are dominated by the insulating planes between bilayers.  

\begin{figure} \includegraphics[width=8.5cm]{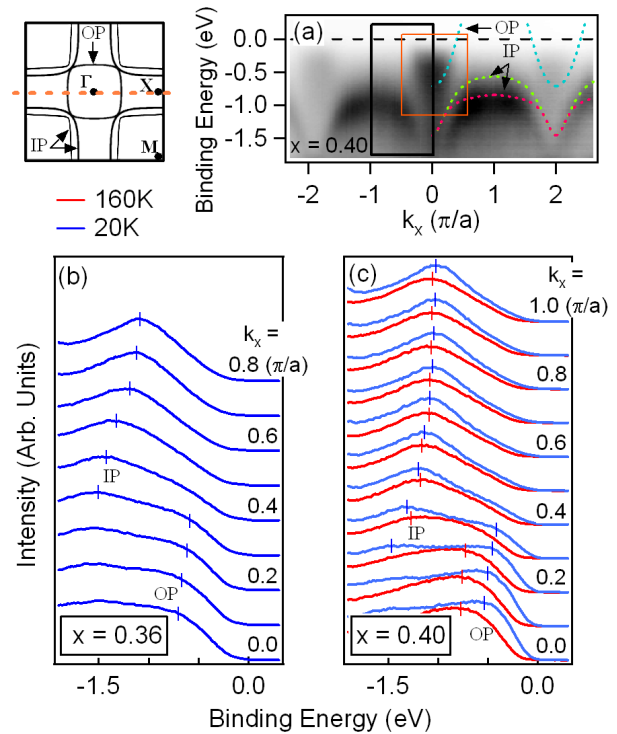}
\caption{\label{fig:GX}(Color online). (a) Photoemission intensity map as a function of binding energy and momentum along cut parallel to the $\Gamma$X direction, defined by the dashed line in the FS inset, for \textit{x}=0.40.  Bandstructure calculations extracted from Ref.~\onlinecite{Huang2000} for the OP and the BB and AB IP bands are overplotted as blue, pink, and green dashed lines, respectively.  (b) EDCs at the labeled momentum positions along this cut for \textit{x}=0.36 at 20K (blue).  High temperature data was not acquired for the full energy range here for \textit{x}=0.36.  Vertical dashes are meant as guides-to-the-eyes of the approximate peak locations of the bands identified as labeled.  The momentum and energy range of these EDCs is illustrated by the black box in panel (a).  The smaller, orange box in panel (a) illustrates the range of data included in Fig.~\ref{fig:OP}.  (c) The corresponding EDCs for \textit{x}=0.40 at 20 (blue) and 160 K (red). 
}\end{figure}

Doping and temperature dependencies of c-axis coupling should also be expressed in the OP band, whose complete \textit{k}-space topology is reported in Figs.~\ref{fig:FS}, \ref{fig:GX}, and \ref{fig:OP}.
The ARPES intensity for a cut along the $\Gamma$X direction, through the electron pocket centered at $\Gamma$ (see orange dashed line in the FS inset), is shown in Fig.~\ref{fig:GX}(a).
Two main sets of bands can be identified in general agreement with the calculated band structure,\cite{Huang2000} overplotted for comparison.
The EDCs corresponding to the area marked by the large black box in panel (a) are shown for \textit{x}=0.36 and \textit{x}=0.40 in Fig.~\ref{fig:GX}(b)~and~\ref{fig:GX}(c), respectively.
For each doping the EDCs contain a clear peak or shoulder at momentum values less than 0.3 ($\pi$/a) related to the OP band, and a single broad peak at larger momentum values related to the IP bands.
As in Fig.~\ref{fig:OGM}, the peaks are too broad to resolve any BLS, do not disperse all the way to \textit{E}$_F$, and the below \textit{T}$_C$ spectral weight reaches closer to $E_F$ near the expected \textit{k}$_F$ than the above \textit{T}$_C$ weight.
Again, these characteristics agree well with previous studies\cite{Dessau1998,Saitoh2000} which discuss them in detail.

To examine the OP band in more detail, Fig.~\ref{fig:OP} shows the ARPES data for the same cut as shown in Fig.~\ref{fig:GX}, focusing on a smaller region near \textit{E}$_F$ [see orange box in Fig.~\ref{fig:GX}(a)].
Similar to the cut in \textit{k}-space studied in Figs.~\ref{fig:OGM} and \ref{fig:BLSX}, although there is a lack of sharp peaks dispersing through \textit{E}$_F$ in the EDCs, there is still important structure in the MDCs.
Figure~\ref{fig:OP}(a) shows a plot of the second derivative (along \textit{k}$_x$ for enhancement of features in momentum distribution) of ARPES intensity for \textit{x}=0.36 at 20K.
This image highlights a clear dispersion which matches well with the calculated OP band dispersion,\cite{Huang2000} overplotted in light blue for \textit{k}$_x\leq0$ and shown in panel (b). 
In the raw data, we resolve dispersing MDC peaks near \textit{E}$_F$ [dark and light blue curves in Fig.~\ref{fig:OP}(c)] corresponding to the OP band, in addition to a broad non-dispersive hump at $k_x=0$, previously solely attributed to the electron pocket,\cite{ChuangCMR2001,Mannella2005} which becomes comparatively dominant at high binding energy.
We note that the corresponding band in the cubic perovskite analogs has been studied with ARPES on thin-film samples,\cite{Shi2004,Chikamatsu2007,Krempasky2008} showing some interesting similarities, however without the complete k-space topology presented here (Fig.~\ref{fig:FS}).
As clear from Fig.~\ref{fig:FS}, the electron pocket is very ``square'', with near-perfect nesting by a vector, \textbf{q}$_N$, parallel to the $\Gamma$X direction.
The size of \textbf{q}$_N$ can be extracted from the \textit{E}$_F$ MDCs as discussed below.

The doping dependence of the OP band shows that its spectral weight is much smaller for \textit{x}=0.40, as reflected in the poorly defined pocket in Fig.~\ref{fig:FS}(c) and weaker MDC peaks compared to the weight at $k_x=0$.
The straight FS contours of the electron pocket suggest that it plays a role in nesting scenarios, in addition to the IP band, where nesting with a vector similar to those reported by scattering experiments \cite{Vasiliu1999,Sun2006} has been previously discussed.\cite{ChuangCMR2001, Sun2006}
From the raw MDCs at \textit{E}$_F$ [Fig.~\ref{fig:OP}(c)], we find that \textbf{q}$_N$ decreases from 0.35 (2$\pi$/a) for \textit{x}=0.36 to 0.31 (2$\pi$/a) for \textit{x}=0.40.
It is interesting to note that the incommensurate scattering vector found to identify short-range polaron correlations follows the opposite trend, increasing from 0.25 (2$\pi$/a) for \textit{x}=0.30 (Ref.~\onlinecite{Kubota2000a}) to 0.30 (2$\pi$/a) for \textit{x}=0.40.\cite{Vasiliu1999}
Our data suggest the polaronic correlations are enhanced at \textit{x}=0.40 (compared to \textit{x}=0.36) due to closer alignment of the cooperating FS nesting instability and polaron correlation vectors, affirming the picture outlined in Fig.~4 of Ref.~\onlinecite{Kim2007}.

\begin{figure} \includegraphics[width=9cm]{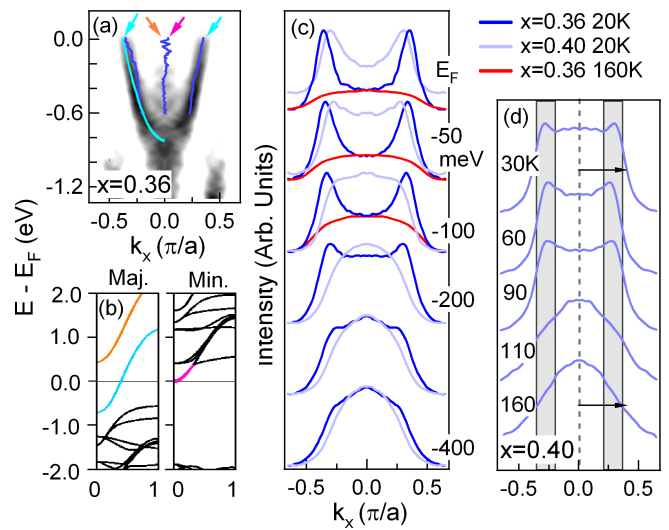}
\caption{\label{fig:OP}(Color online).  (a) Second derivative ($d^2/dk^2$) of ARPES intensity map for \textit{x}=0.36 at 20K, along $\Gamma$X [data range highlighted by small orange box in Fig.~\ref{fig:GX}(a)], integrated +/- 1/8 $\pi/a$ along \textit{k}$_y$, and symmetrized around \textit{k}$_x$=0.  The unsymmetrized MDC peak dispersion (using a three-Lorentzian fit) is plotted in dark blue.  The light blue curve is the dispersion from panel (b).  (b) Bandstructure calculations extracted from Ref.~\onlinecite{Huang2000}.  The light blue (orange) bands are the BB (AB) band from the BLS of the majority 3\textit{d}$_{3z^2-r^2}$ band and the pink band is a minority band.  (c) Symmetrized MDCs (normalized to equal peak intensity after a constant background subtraction) below T$_C$ for x=0.36 (dark blue), \textit{x}=0.40 (light blue), and above \textit{T}$_C$ for \textit{x}=0.36 (red).
(d) Normalized and symmetrized MDCs at E$_F$ for \textit{x}=0.40 at several temperatures (collected from high to low \textit{T}, ensuring that the high \textit{T} lack of features is not due to surface degradation).  The gray stripe marks the peak location below \textit{T}$_C$, and the vertical dotted line marks the spectral weight at $\Gamma$.
The disappearence of the peaks above \textit{T}$_C$ is not due to broadening of the central hump as the width of the combined peak and hump structure (shown by the black arrow at 30 K) is larger than the remaining hump at 160 K (same arrow copied for comparison at 160 K).
}\end{figure}

The temperature dependence of the OP band is shown in Figs.~\ref{fig:OP}(c)~and~\ref{fig:OP}(d).
For \textit{x}=0.36, the MDC peaks below \textit{T}$_C$ [dark blue curves in Fig.~\ref{fig:OP}(c)] fade into possible shoulders of the central hump above \textit{T}$_C$ (red curves).
The MDC peaks for \textit{x}=0.40 behave similarly.
Figure~\ref{fig:OP}(d) displays the \textit{x}=0.40 MDC at \textit{E}$_F$ for several temperatures, showing the peaks disappear right before \textit{T}$_C$ $\sim$~120K.
These results fit the picture of a``pseudogap'' smearing spectral weight away from \textit{E}$_F$ above \textit{T}$_C$,\cite{Saitoh2000} accompanied by a loss of coherence.\cite{Mannella2005,Mannella2007}  
Our data show that this effect also occurs in the OP band at both dopings.
These temperature and doping dependencies are in line with the above scenario of further c-axis confinement (within the bilayer) associated with the FM to PI transition and the FM to antiferromagnetic canting transition, respectively.

At last we discuss possible origins for the hump at $k_x$=0.
Although further studies are needed to sort out the nature of this spectral feature, a couple possible scenarios can be invoked from bandstructure calculations: (1) it results from a shallow, or even unoccupied, minority band close to \textit{E}$_F$ at $\Gamma$ [pink in Fig.~\ref{fig:OP}(b)]; and (2) it is a signature of the unoccupied AB OP band, shown in orange in Fig.~\ref{fig:OP}(b) (the OP band should have BLS like the IP band, only in this case the AB is split above \textit{E}$_F$).  
This second scenario can account for the decrease (from \textit{x}=0.36 to 0.40) in the ratio of MDC peak intensity at \textit{k}$_x>0$ to that of the hump at \textit{k}$_x$=0, as the doping dependent decrease of BLS discussed above may result in a shift of the `orange' band down in energy, giving stronger intensity to the hump at $\Gamma$.
The weight may also be related to slight \textit{k}$_z$ broadening, due to a finite amount of \textit{k}$_z$ dispersion expected in the electron pocket.\cite{Huang2000}
A similar explanation is given in Ref.~\onlinecite{Krempasky2008} for similar spectral weight found in the perovskite samples.

\section{\label{sec:Summary}SUMMARY}

In conclusion, by using higher photon energy and out-of-plane polarization with respect to previous studies, we reported the full FS topology resolving complete bilayer splitting of the in-plane, 3\textit{d}$_{x^2-y^2}$-derived band at \textit{x}=0.36 and the out-of-plane, 3\textit{d}$_{3z^2-r^2}$-derived band at both \textit{x}=0.36 and \textit{x}=0.40.
The peculiar doping and temperature dependencies provide direct evidence of a change in the electronic freedom along the c-axis between planes within the bilayer, a unique aspect of the electronic structure not observeable by bulk measurements.
These results provide a possible explanation for the differences recently reported for these two dopings and suggests an interesting role for dimensionality in the CMR transition.

\begin{acknowledgments}
We would like to thank D.-H. Lee and E. Artacho for very useful discussions.
The ARPES work was supported by the Director, Office of Science, Office of Basic Energy Sciences, Materials Sciences and Engineering Division, of the U.S. Department of Energy under Contract No. DE-AC02-05CH11231.
The ARPES work was performed at the Advanced Light Source, Lawrence Berkeley National Laboratory, which is supported by the Director, Office of Science, Office of Basic Energy Sciences, of the U.S. Department of Energy under Contract No. DE-AC02-05CH11231
Sample growth was performed at Argonne National Laboratory and is supported by the US DOE Office of Science-Basic Energy Sciences, Division of Materials Science under Contract No. DE-AC02-06CH11357.
\end{acknowledgments}

\bibliographystyle{apsrev}

\end{document}